\documentclass[aps,pre,preprint,superscriptaddress]{revtex4}

\usepackage{graphicx,latexsym}

\begin{document}


\title{Effective slip boundary conditions for flows over nanoscale chemical heterogeneities}


\author{S. C. Hendy}
\affiliation{MacDiarmid Institute for Advanced Materials
and Nanotechnology, Industrial Research Ltd, Lower Hutt, New Zealand}
\affiliation{MacDiarmid Institute of Advanced Materials and Nanotechnology,
School of Chemical and Physical Sciences, Victoria University of Wellington, PO Box 600, Wellington, New Zealand}
\author{N. J. Lund}
\affiliation{MacDiarmid Institute of Advanced Materials and Nanotechnology,
School of Chemical and Physical Sciences, Victoria University of Wellington, PO Box 600, Wellington, New Zealand}

\email[]{s.hendy@irl.cri.nz} 
\homepage[]{http://www.victoria.ac.nz/scps/research/compnanotech/}


\date{\today}

\begin{abstract}
We study slip boundary conditions for simple fluids at surfaces with nanoscale chemical heterogeneities. Using a perturbative approach, we examine the flow of a Newtonian fluid far from a surface described by a heterogeneous Navier slip boundary condition. In the far-field, we obtain expressions for an effective slip boundary condition in certain limiting cases. These expressions are compared to numerical solutions which show they work well when applied in the appropriate limits. The implications for experimental measurements and for the design of surfaces that exhibit large slip lengths are discussed.  
\end{abstract}


\maketitle


\section{Introduction}

The no-slip boundary condition was considered to have been experimentally established for simple liquids in the early 20th century.  However, the refinement of a number of measurement techniques has recently led to the observation of nanoscale, and even micron-scale, violations of the no-slip boundary condition by simple fluids flowing over non-wetting surfaces \cite{Zhu01}. In many instances however, poorly controlled microscopic factors that influence the measured macroscopic slip length, such as roughness, chemical heterogeneity, or contaminants such as air bubbles, have lead to  apparent discrepancies in the magnitude of slip reported in the literature \cite{Cecile05,Neto05,Lauga07}. Thus, it is important to distinguish between {\em effective} or apparent slip, typically measured in macroscopic experiments, which emerges from the interaction of microscopic chemical heterogeneity, roughness and contaminants, and {\em intrinsic} slip, which arises solely from the chemical interaction between the liquid and a homogeneous, atomically flat surface.

Slip is usually described in fluid mechanics by the Navier slip boundary condition \cite{Navier}. This states that at a solid boundary, $z=0$, the slip velocity, $u$, is proportional to the shear rate, $\partial_z u$ i.e.
\begin{equation}
\delta \left. \partial_z u \right|_{z=0} = \left. u \right|_{z=0}
\end{equation}
where the constant of proprtionality $\delta$ is called the slip length. In some instances, experiments have found that the slip length can range from nanometers \cite{Joly06} to tens of micrometers \cite{Choi06}. As slip on this scale can profoundly affect flows in micro and nanofluidic devices, these findings have generated considerable interest \cite{Granick03}. For instance, large effective slip lengths potentially offer new ways of controlling flows in microdevices \cite{Hendy05, Rothstein07}. From a theoretical point of view, neither intrinsic nor effective slip lengths can yet be predicted microscopically. Nonetheless, a useful way to study intrinsic slip is through atomistic computer simulation, using techniques such as molecular dynamics. Such studies suggest that flows over flat hydrophilic surfaces will exhibit intrinsic slip lengths less than a few nanometers, while flows over flat hydrophobic surfaces should have slip lengths of tens of nanometers \cite{Barrat99}. Indeed, strong experimental support for this picture is now beginning to emerge \cite{Joly06}. 

\begin{figure}
\resizebox{\columnwidth}{!}{\includegraphics{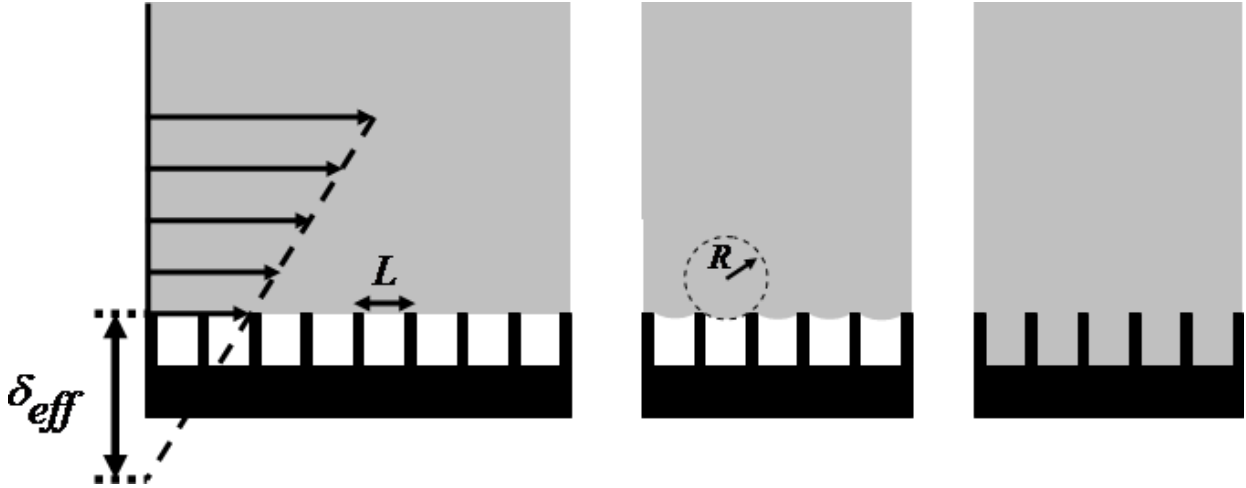}}
\caption{\label{surface-diagram} The figure illustrates the flow over a highly heterogeneous hydrophobic surface characterized by length scale $L$. At low pressures, the liquid is in the Cassie state, where it does not penetrate into the surface, leading to large effective slip lengths (left). At intermediate pressures the liquid begins to penetrate the surface (described by radius of curvature $R$) and the effective slip decreases (center). Finally, at sufficiently high pressures the liquid penetrates the surfaces, which will drastically reduce the effective slip length (right).}
\end{figure}

However, it is on so-called superhydrophobic surfaces that slip lengths as large as tens of micrometers have been observed \cite{Choi06,Joseph06} (see figure~\ref{surface-diagram}). The best known example of a superhydrophobic surface is the leaf of the lotus plant, which possesses a microstructure and surface chemistry that prevents water from wetting its surface, leading to droplet contact angles close to 180$^o$ \cite{Barthlott}. Recently, nanotechnologists have learnt to mimic this so-called Lotus effect by creating superhydrophobic surfaces \cite{Cheng05} using materials such as carbon nanotubes assembled in dense forests \cite{Lau03}. The repulsion of water by such surfaces means that droplets or larger scale flows are essentially lubricated by a layer of air, leading to what is clearly a large effective slip length, with drag only occurring at the few points of the surface where the flow makes contact with the substrate. Again there is no rigorous theoretical description of how such effective slip lengths depend on the underlying microstructure of such highly heterogeneous surfaces. 

It is of interest then to study how effective slip lengths emerge from heterogeneous intrinsic slip lengths. Such problems have been studied by numerical methods including molecular dynamics \cite{Cecile03}, lattice Boltzmann simulations \cite{Toschi06} and numerical solutions of the Stokes equations \cite{Cecile04}. In addition, some exact solutions are known for flow in channels both over and along stripes of alternating no-slip ($\delta = 0$) and perfect slip ($\delta = \infty$) \cite{Philip72,Lauga03,Prosperetti07}. However we are still unable to deduce effective slip lengths from the underlying microstructure and chemistry of a surface in the general case. A deeper understanding of effective slip would give insight into how artifacts such as roughness and nanobubbles effect experimental measurements of slip. It may also allow optimization of superhydrophobic surfaces to extremize slip lengths for use in devices \cite{Hendy05,Rothstein07}. 

In this paper our goal is to calculate effective, far-field slip lengths on chemically heterogeneous surfaces, including nanoporous surfaces and surfaces covered in nanobubbles, which possess finite slip lengths $0 < \delta < \infty$. In contrast, previous work has generally focused on the case where $\delta$ is 0 or $\infty$. We begin by considering the hierarchy of length scales present in the problem and use this to define several distinct sets of problems that arise from this hierarchy in limiting cases. We then derive approximate solutions to two these problems to arrive at expressions for the effective slip length in the corresponding limiting cases. Finally, we discuss the implications of these expressions for experiments and for the design of surfaces with large effective slip lengths.    

\section{Analysis}

The general problem we will examine here concerns the Stokes flow of an incompressible fluid past a surface described by a finite slip length that is a function of position on the surface. We will consider simple shear flows, as might be encountered in many experimental situations. Thus at some distance $W$ away from the heterogeneous surfaces, we apply a shear in the $x$-direction either at constant rate $\dot{\gamma}$ or with a constant velocity $u=u_s$ in the $x$-direction. For the constant shear rate, this results in the following boundary conditions at $z=W$ : $\,\left.\partial_z u \right|_{z=W} = \dot{\gamma}$ where $u$ is the $x$-component of the velocity $\vec{u}$. For a shear flow past a flat surface which lies in the $xy$ plane, $z=0$, the slip length $\delta = \delta(x,y)$ leads to the boundary condition $\delta(x,y) \left. \partial_z u \right|_{z=0} = u|_{z=0}$. Note that we will initially ignore the effects of roughness induced by curvature of the liquid-vapor interface \cite{Cecile04} as shown in figure~\ref{surface-diagram}. This amounts to assuming that the radius of curvature of the interface $R$ is much larger than the other length scales in the problem. The effective slip length is then given in the limit as $W \rightarrow \infty$ by the expression
\begin{equation}
\delta_{eff} \left. \partial_z u \right|_{z=\infty} = u|_{z=\infty}. 
\end{equation}

There are three cases we have considered: $\delta=\delta(y)$ where $\delta(y)$ is periodic with period $L$ (i.e. the shear is parallel to the patterning), $\delta=\delta(x)$ where $\delta(x)$ is periodic with period $L$ (i.e. the shear is perpendicular to the patterning) and $\delta=\delta(x,y)$ where $\delta(x,y)$ is periodic in the $x$ and $y$ direction with unit cell $(L_x,L_y)$ (i.e. the flow occurs over rectangular patches). We will focus on slip lengths which are patterned in stripes or patches with sharp edges so that $\delta(x,y)$ will generally be considered to be a piecewise constant function. We will also assume that $0 < \delta(x,y) < \infty$. For instance, in the case where the stripes oriented parallel to the direction of shear:
\[ \delta(y) = \left\{ \begin{array}{lc}
               \delta_1 & 0 \leq y \leq a         \\
               \delta_2 & a < y \leq L
          \end{array} \right. \]
where $a$ is the stripe width and $\delta_1 < \delta_2$. In figure~\ref{surface-diagram}, $\delta_1$ would be the slip length of the solid surface, $\delta_s$ and $\delta_2$ would be the slip length over the vapor regions, $\delta_g$. In what follows we will restrict ourselves to presenting the analysis for the first geometry, where $\delta=\delta(y)$, and simply report the analogous results of our calculations for the other two cases. Although the analysis is simplest for this first case, the approach in latter two geometries does not differ significantly from that presented here.   

It is useful at this stage to consider the magnitudes of the relevant length scales in the problem. Here we will assume that intrinsic slip lengths for smooth solid surfaces are at most 10-20 nm, consistent both with recent measurements of the the slip lengths for hydrophobic surfaces \cite{Joly06} and with the results of molecular dynamics simulations \cite{Barrat99}. To estimate slip lengths at the liquid-vapor interface, we will use De Gennes's expression \cite{deGennes}: $\delta_g \sim (\mu_l/\mu_g) t$ where $\mu_{l(g)}$ is the viscosity of the liquid (gas) and $t$ is the thickness of the gas layer. For pure water flowing over air at room temperature, we estimate that $\delta_g \sim 50 t$. The length scales that describe the patterning can clearly vary widely. For the thiol functionalized carbon nanotube forests studied in Ref \cite{Joseph06}, $L \sim 1$ $\mu$m, $a \sim 50-100$ nm and $t > 1$ $\mu$m, giving the following hierarchy of lengths: $\delta_s \ll a \ll L \ll \delta_g$. The mixing device constructed in Ref~\cite{Rothstein07} has grooves of width $L-a \sim 100$ $\mu$m and depth $t \sim 50$ $\mu$m, spaced at $L \sim 10$ mm, giving $\delta_s \ll (L-a) \ll \delta_g \sim L$. In contrast, a hydrophobic surface contaminated by nanometer-sized bubbles or a hydrophobic nanoporous surface might be described by the hierarchy: $a \sim L \ll \delta_s$ and $L \ll \delta_g$. The nanostructured hydrophobic channels studied by molecular dynamics simulations in Refs. \cite{Cecile03,Cecile04} would also be likely to satisfy this hierarchy. More typically however, unless it has been specially prepared or contaminated by vapor bubbles, we might expect a heterogeneous surface to be described by $\delta_1 \sim \delta_2 \ll a \sim L$. 

The hierarchy of length scales present will effect the magnitude of the terms in the slip boundary condition. Introducing non-dimensional spatial coordinates $\hat{x}=x/L$, $\hat{y}=y/L$ and $\hat{z}=z/L$, the boundary condition for stripes oriented parallel to the direction of shear:
\begin{eqnarray}
\left(\delta_1 / L \right) \left. \partial_{\hat{z}} u \right|_{\hat{z}=0} =  u|_{\hat{z}=0}, & 0 < \hat{y} \leq \beta  \nonumber \\ 
               \left(\delta_2 / L \right) \left. \partial_{\hat{z}} u \right|_{\hat{z}=0} =  u|_{\hat{z}=0}, & \beta < \hat{y} \leq 1,
\end{eqnarray}
where $\beta = a/L$. In what we expect to be the most common situation, where $\delta_1$ and $\delta_2 \ll  L$, the boundary conditions are no-slip to zeroth order:  
\begin{eqnarray}
               u|_{\hat{z}=0} = O\left(\delta_1 / L\right),   & 0 < \hat{y} \leq \beta   \nonumber      \\
               \label{case-one} u|_{\hat{z}=0} = O\left(\delta_2 / L\right),  & \beta < \hat{y} \leq 1.
\end{eqnarray} 
For hydrophobic surface covered by nanometer scale bubbles, or a nanoporous substrate with $L < 10$ nm, we would have $L \ll \delta_s$ and $L \ll \delta_g$, so that the boundary conditions are shear free at zeroth order:
\begin{eqnarray}
               \left. \partial_{\hat{z}} u \right|_{\hat{z}=0} = O\left(L / \delta_s \right), & 0 < \hat{y} \leq \beta  \nonumber \\
               \label{case-two} \left. \partial_{\hat{z}} u \right|_{\hat{z}=0} = O\left(L / \delta_g \right),  & \beta < \hat{y} \leq 1.
\end{eqnarray}  
For superhydrophobic surfaces, such as those fabricated in Ref.~\cite{Joseph06}, which satisfy $\delta_s \ll L \ll \delta_g$, the boundary conditions are mixed at zeroth order: 
\begin{eqnarray}
               u|_{\hat{z}=0} = O\left(\delta_s / L \right),   & 0 < \hat{y} \leq \beta     \nonumber    \\
               \label{case-three} \left. \partial_{\hat{z}} u \right|_{\hat{z}=0} =  O\left( L / \delta_g\right), & \beta < \hat{y} \leq 1.
\end{eqnarray}
Surfaces can be similarly defined both in the case of stripes perpendicular to the shear $\delta=\delta(x)$, and in the case of patches or more complex patterns where $\delta=\delta(x,y)$. In these cases we consider $\beta$ to be the area fraction of the solid (or more generally the area fraction of the surface with the smaller slip length $\delta_1$).

Our approach is to treat these problems perturbatively, solving them to first order in the relevant small parameters. As we will show below, this perturbative approach succeeds for boundary conditions (\ref{case-one}) and (\ref{case-two}) but fails for boundary condition (\ref{case-three}). In addition, for surfaces that satisfy (\ref{case-three}), exact solutions of the Stokes equations with the zeroth order boundary condition are only known in the case where the shear is parallel or perpendicular to the stripes \cite{Philip72}. 

In fact, when the shear is parallel to the stripes ($\delta = \delta(y)$), as both the velocity and the pressure are a function of $y$ and $z$ only, the equation for the $x$-component of the velocity, $u$, is just Laplace's equation:  
\begin{equation}
\label{Laplace}
\partial^2_{\hat{y}} u + \partial^2_{\hat{z}} u = 0. 
\end{equation} Indeed this is the easiest geometry to treat, and as such we will generally restrict ourselves to presenting the analysis of this particular case. Nonetheless, apart from case (\ref{case-three}), we have been able to extend our calculations of effective slip lengths to the more general patterning where $\delta = \delta(x,y)$.    

We now seek a solution of (\ref{Laplace}) that satisifies the boundary condition (\ref{case-two}) of the form:
\begin{equation}
\label{expansion}
u(\hat{x},\hat{z}) = u_0(\hat{y},\hat{z}) + \alpha u_1(\hat{y},\hat{z}) + O\left(\alpha^2\right) 
\end{equation}   
i.e. an asymptotic series in $\alpha = L/\delta_1$. As equation~(\ref{Laplace}) is linear, each of the terms in the expansion, $u_i$, are solutions of (\ref{Laplace}). The $\hat{z}=0$ boundary condition (\ref{case-two}) at zeroth order in $\alpha$ is just a shear-free condition: 
\begin{equation}
\frac{\partial u_0}{\partial \hat{z}}(\hat{y},0) = 0 
\end{equation}
for $0 < \hat{y} < 1$ and the solution at this order is simply a homogeneous shear-free flow with $u_0=u_s$. At first order in $\alpha$ the slip boundary condition becomes
\begin{eqnarray}
\partial_{\hat{z}} u_1 (\hat{x},0) = u_s  & 0 < \hat{y} \leq \beta \nonumber \\
\partial_{\hat{z}} u_1 (\hat{x},0) = \frac{\delta_1}{\delta_2} u_s  & \beta < \hat{y} \leq 1.
\end{eqnarray}  
Further conditions follow from the lack of a pressure head, the periodicity of the flow in the $y$-direction (period $L$) and the fact that the velocity
component of the flow normal to the surfaces at $z=0$ and $z=\infty$ must vanish.

To solve the first-order problem, we use the periodicity of the flow in the $\hat{y}$-direction to write $u_1$ as a Fourier series as follows:
\begin{equation}
\label{FourierSeries}
u_1(\hat{y},\hat{z}) = \sum^{\infty}_{n=0} U_n\left(\hat{z}\right) \exp \left( i k_n \hat{y}\right) 
\end{equation}
where $k_n = 2 \pi n$ and 
\begin{equation}
U_n(\hat{z}) = \int^{1}_{0} u_1\left(\hat{y},\hat{z}\right) \exp \left( i k_n \hat{y} \right) d\hat{y}. 
\end{equation}
Inserting (\ref{FourierSeries}) into (\ref{Laplace}) we find that 
\begin{equation}
U_0 \left( \hat{z} \right) = A_0 + B_0 \hat{z}  
\end{equation} 
and
\begin{equation}
U_n \left( \hat{z} \right) = A_n e^{-k_n \hat{z}} + B_n e^{k_n \hat{z}},
\end{equation} 
for $n>0$. Furthermore, for $n>0$ as the upper boundary conditions on $U_n \left( \hat{z} \right)$ apply at $\hat{z} = W/L \gg 1$, the coefficients $B_n$ will be of order $\exp \left( - k_n W / L \right)$, so we may neglect these in the far field. 

The slip condition at $\hat{z}=0$ gives  
\begin{eqnarray}
B_0  & = & \frac{d U_0}{d \hat{z}}\left(0\right) = \int^{1}_{0} \frac{\partial u_1}{\partial \hat{z}} \left(\hat{y},0\right) d\hat{y} \nonumber \\
& = & u_s \left( \beta + (1-\beta)\frac{\delta_1}{\delta_2} \right)
\end{eqnarray} 
so that 
\begin{equation}
U_0 \left( \hat{z} \right) = u_s  \left( \hat{z} - W/L \right) \left( \beta + (1-\beta)\frac{\delta_1}{\delta_2} \right). 
\end{equation} 
Thus the solution to the first-order problem is given by 
\begin{equation}
u_1 \left( \hat{y}, \hat{z} \right) = u_s \left( \hat{z} - W/L \right) \left( \beta + (1-\beta)\frac{\delta_1}{\delta_2} \right) + \sum^{\infty}_{n=1} A_n e^{- k_n \hat{z}}.  
\end{equation} 
It is easily verified then that to first order in $\delta_s/L$, as $u \left( \hat{x}, \hat{z} \right) = u_s + \alpha u_1 +O \left( \alpha^2 \right)$, the effective slip length for flow over parallel stripes is given by
\begin{equation}
\label{deltaeff} 
\frac{1}{\delta_{eff}} = \frac{\beta}{\delta_1} + \frac{1-\beta}{\delta_2}. 
\end{equation} 

The analysis in the remaining two geometries is similar ($\delta=\delta(x)$ and $\delta = \delta(x,y)$), although somewhat more complicated as the Stokes equations do not reduce to the Laplace equation (equation (\ref{Laplace})) in these cases. Nonetheless, both in the case of shear directed perpendicular to stripes of fractional width $\beta$ and in the case of regular patches of area fraction $\beta$, we again find that this relation holds. We note that these relationships have previously been observed to hold empirically for numerical solutions of the steady state Stokes equations \cite{Cecile04}. In the second case, (\ref{case-two}), where $\delta_1$ and $\delta_2$ are much less than $L$, a similar analysis to that given above reveals that $\delta_{eff} = \beta \delta_1  + \left( 1-\beta \right) \delta_2$. Furthermore, if $\delta_1 \sim \delta_2 \gg L$, equation (\ref{deltaeff}) also reduces to $\delta_{eff} \simeq \beta \delta_1  + \left( 1-\beta \right) \delta_2$.   

\begin{figure}
\resizebox{\columnwidth}{!}{\includegraphics{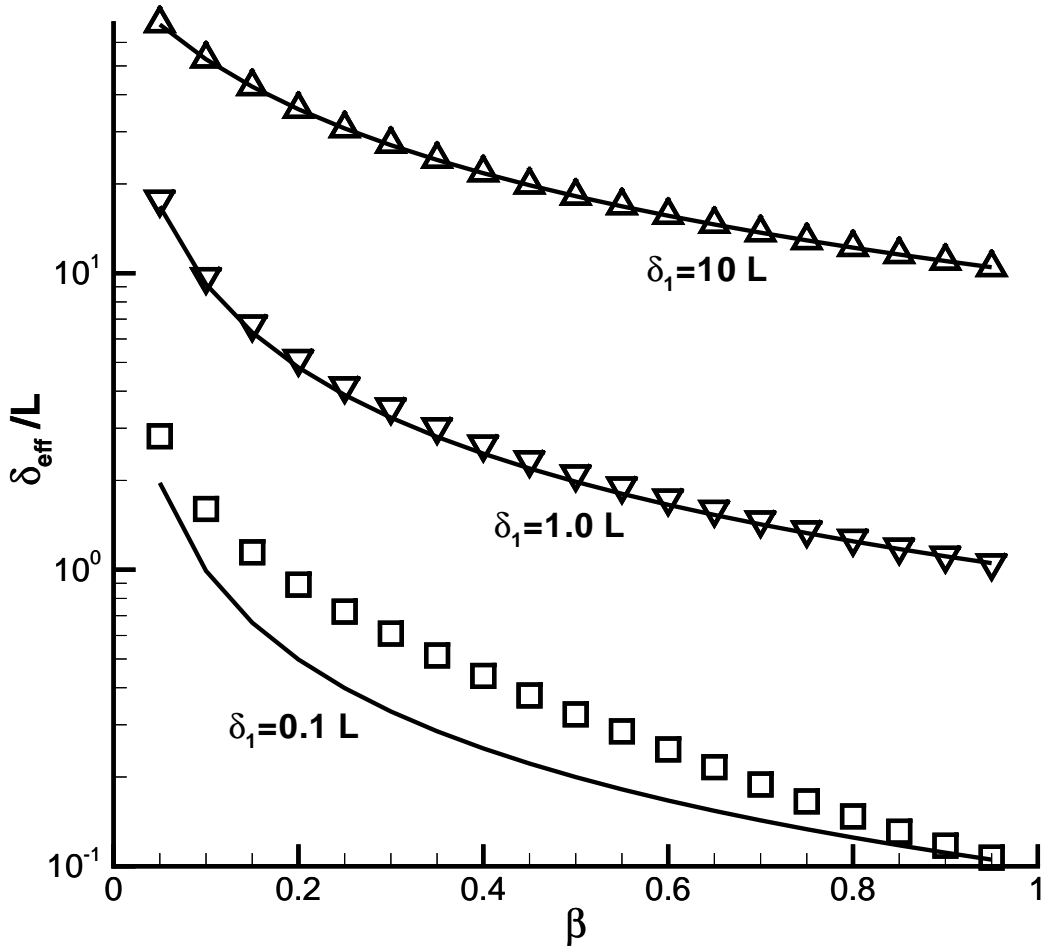}}
\caption{\label{Fig2} Effective slip length as a function of stripe width $\beta$ as given by equation (\ref{deltaeff}) (solid line) and by finite difference solutions of the Laplace equation (symbols) for the case of flow over parallel stripes for $\delta_1 = 0.1 - 10 L$ and $\delta_2 = 10^2 L$.}
\end{figure}
\begin{figure}
\resizebox{\columnwidth}{!}{\includegraphics{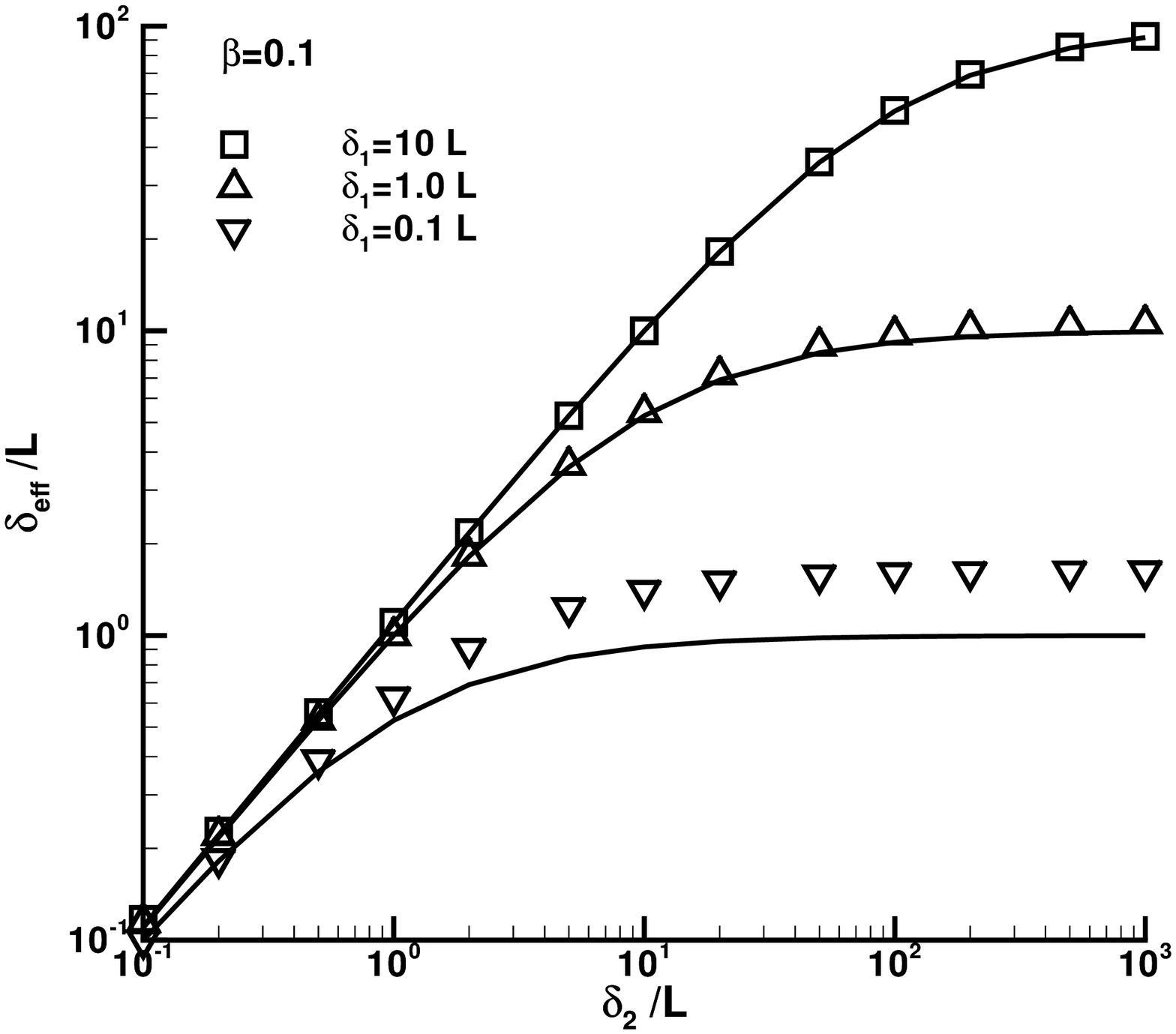}}
\caption{\label{Fig3} Effective slip length for $\beta=0.1$ as a function of $\delta_2$ for several values of $\delta_1$ as given by equation (\ref{deltaeff}) (sold line) and by finite difference solutions of the Laplace equation (symbols) for the case of flow over parallel stripes.}
\end{figure}

These expressions can be tested numerically. Figure~\ref{Fig2} compares the effective slip length inferred from by finite difference solutions of the Laplace equation for the case of flow directed along stripes ($\delta = \delta(\hat{y})$) with equation (\ref{deltaeff}) as a function of stripe width $\beta$ for $\delta_1 = 0.1 - 10 L$ and $\delta_2 = 10^2 L$. The plot shows that equation (\ref{deltaeff}) gives an excellent approximation to the effective slip length for $\delta_1 \geq 1.0 L$ but underestimates the effective slip length by up to a factor of 2 for $\delta_1 = 0.1 L$ while still tracking the dependence of the slip length on $\beta$. Similarly, Figure~\ref{Fig3} shows the effective slip length as a function of $\delta_2$ for $\delta_1 = 0.1 - 10 L$ with $\beta = 0.1$ for flow parallel to stripes. Again it is apparent that equation (\ref{deltaeff}) is very successful for $\delta_1 \geq L$ but underestimates $\delta_{eff}$ for $\delta_1 = 0.1 L$ as one might expect. 

Unfortunately, as noted above, the perturbative approach used here fails for the third case (\ref{case-three}), where $\delta_s \ll L \ll \delta_g$. The exact solution to the zeroth order problem is only known for the parallel stripes case ($\delta = \delta(y)$), and this solution is not differentiable at the heterogeneous surface \cite{Philip72} causing the expansion (\ref{expansion}) to fail. At zeroth order (in $\delta_s/L$ and $L/\delta_g$), for the parallel stripes, the effective slip length is given by $\delta_{eff}^0 = \frac{L}{\pi} \ln{\left( \csc{\beta \pi/2}\right)}$ \cite{Philip72}. Note that there are also solutions known to this problem in the transverse stripe case ($\delta=\delta(x)$) \cite{Philip72}, which differs by a factor of one half from the perpendicular case, but not to the more general case ($\delta=\delta(x,y)$).

Finally, in this section, we note that (\ref{deltaeff}) can be generalized to more complex patternings $\delta=\delta(x,y)$. Provided $\delta(x,y) \gg L$ everywhere in the domain then a similar analysis to that given above yields 
\begin{equation}
\label{deltagen} 
\delta_{eff} = \langle \frac{1}{\delta(x,y)} \rangle^{-1} 
\end{equation}
where the angle-brackets denote that mean value of the function over the surface. Similarly, if $\delta(x,y) \ll L$ everywhere on the surface $\delta_{eff} = \langle \delta(x,y) \rangle$. Thus the results presented here are not restricted to sharp, well-defined patternings. 

\section{Discussion}

The implications of (\ref{deltaeff}) for the effective slip lengths of hydrophobic surfaces contaminated by nanobubbles induced by roughness such as the superhydrophobic surfaces studied in Ref.~\cite{Cecile04} are interesting. Here, if $a \sim t \sim L < 10$ nm for instance, we would expect $\delta_g \sim$ 50 $t \gg \delta_s \sim$ 10-20 nm $> L$, so that according to our results here, $\delta_{eff} \sim \delta_s/\beta$. Thus the apparent increase in slip length for a hydrophobic surface contaminated by nanoscale bubbles remains proportional to $\delta_s$ and will be effectively independent of $\delta_g$. A surface with effective slip described by (\ref{deltaeff}) could exhibit slip lengths several times larger than $\delta_s$ e.g. for $a/L = 0.5$, $\beta = 0.25$ so that $\delta_{eff}$ could be as large as 40-80 nm. This is certainly consistent with many experimental measurements of slip (e.g. see the review Ref.~\cite{Lauga07}).  

It is important to note however that we have not considered the effects of roughness in our calculations above. Following Ref.~\cite{Panzer90}, one can consider the Navier slip boundary condition for fluid flow past a general solid surface:
\begin{equation}
\left. {\bf t} \cdot {\bf u} \right|_{z=z(x,y)} = \left. \delta t_i n_j \left( \nabla_i u_j + \nabla_j u_i \right) \right|_{z=z(x,y)}  
\end{equation}
where ${\bf t}$ and ${\bf n}$ are the tangent and normal to the surface respectively. This can be rewritten in terms of the radius of curvature of the surface, $R$, at each point on the surface as follows:
\begin{equation}
\label{Panzer}
{\bf t} \cdot {\bf u} |_{z=z(x,y)} = \frac{\delta}{\left( 1 - \delta/R \right)} \, {\bf n} \cdot {\bf \nabla} \left( {\bf t} \cdot {\bf u} \right) |_{z=z(x,y)} 
\end{equation}
as originally noted in Ref.~\cite{Panzer90}. Thus, by using (\ref{Panzer}) to incorporate the effects of curvature of the liquid-vapor interface on the slip length at this surface, the effective slip length can be written 
\begin{equation} 
\delta_{eff} = \left( \frac{\beta}{\delta_s} +\left(1-\beta\right)\left(\frac{1}{\delta_g} - \frac{1}{R}\right) \right)^{-1}. 
\end{equation}
From this expression it can be seen that the curvature will become important when $R \sim \delta_s$ and will decrease the effective slip length. 

As noted above, when $\delta_s < L$, equation (\ref{deltaeff}) fails as can be seen in the numerical calculation (figures~\ref{Fig2} and \ref{Fig3}). For parallel or perpendicular stripes, when $\delta_s \ll L \ll \delta_g$, we expect that $\delta_{eff}$ should be given by the expressions due to Philip \cite{Philip72}: that is $\delta_{eff} \sim L \ln \left( \csc \beta \pi/2 \right)$. Although neither exact nor approximate analytic solutions are known for the two-dimensional case, experiments \cite{Joseph06} and numerical solutions \cite{Ybert07} suggest that $\delta_{eff} \sim L$ also for fixed $\beta$ in this limit. Thus for fixed $\beta$ there appears to be a cross-over from $\delta_{eff} \sim \delta_s$ to $\delta_{eff} \sim L$ as $\delta_s$ goes from above to below $L$. We note that Ybert et al. \cite{Ybert07} have suggested that the expression:
\begin{equation} 
\label{Ybert}
\delta_{eff} \sim \frac{\delta_s+a}{\beta} 
\end{equation}
may approximately interpolate between these two limits. This is consistent with (\ref{case-three}), which would lead one to expect a correction to the zeroth order solution due to the finite slip length of the solid proportional to $\delta_s$.

Finally, these results suggest that very large slip lengths ($> 100$'s of nanometers) cannot be achieved by structuring a hydrophobic surface on length scales of 10's of nanometers as $\delta_{eff} \sim \delta_s$ in this case. However, it suggests that a hierarchy of length scales, which can lead to considerable enhancements in contact angle, could also be a useful way of maximizing effective slip length. If a hydrophobic substrate were patterned both on nanometer length scales $L^\prime < \delta_s$ {\em and} on micrometer length scales $L$, one might enhance $\delta^\prime_{eff}$ for the solid by a factor of 3-4 as discussed above. According to equation (\ref{Ybert}), this could lead to a substantial increase in the overall effective slip length $\delta_{eff}$ if $\delta^\prime_{eff} \sim a$ or larger.     

\section{Conclusion}

In summary, we have considered surfaces with alternating stripes or patches of slip length $\delta_1$ and $\delta_2$ patterned on a length scale $L$. In the far-field, we derived expressions for the effective or apparent slip length in several cases. When $\delta_1$ and $\delta_2$ $\ll L$ or when $\delta_2 \sim \delta_1 \gg L$, the effective slip length is the area weighted average of the two slip lengths: $\delta_{eff} = \beta \delta_1 + (1-\beta) \delta_2$ where $\beta$ is the area fraction of slip length $\delta_1$. When $\delta_2 \gg \delta_1$ $\gg L$, the effective slip length is given by $1/\delta_{eff} = \beta / \delta_1 + (1-\beta) / \delta_2$. These expressions have previously been found to hold empirically in molecular dynamics and other numerical simulations of flows over nanostructured superhydrophobic surfaces \cite{Cecile04}. The derivation provided here now provides theoretical support for these relationships and elucidates their range of validity.

\begin{acknowledgments}
The authors would like acknowledge useful discussions on this topic with Cecile Cottin-Bizonne, Catherine Barentin and Christophe Ybert. The authors also acknowledge partial support through the New Zealand Foundation for Research, Science and Technology contract number CO8X0409.  
\end{acknowledgments}

\end{document}